\documentclass[11pt]{article} 
\newcommand{\ignore}[1]{\index{ignore}}
\usepackage{etex} 
\usepackage{libertine}

\usepackage{sfmath}
\usepackage{graphicx,soul,comment,booktabs} 
\usepackage[table]{xcolor}
\usepackage[top=2cm,bottom=2.5cm,left=1.5cm,right=1.5cm]{geometry}
\usepackage{amsmath,amssymb,amsfonts,bm,framed,authblk,float,epstopdf}
\usepackage[colorlinks=true,linkcolor=dblue,citecolor=dred]{hyperref}
\usepackage[nameinlink]{cleveref}
\usepackage[labelfont=bf,labelsep=period]{caption}
\usepackage[sort&compress,authoryear,comma]{natbib}
\bibpunct{(}{)}{;}{a}{,}{,}
\def\cite#1{\citep{#1}}

\usepackage{xr} 

\usepackage{multicol} 

\usepackage{caption}
\definecolor{dgreen}{rgb}{0,0.6,0} \definecolor{dred}{rgb}{0.6,0,0}
\definecolor{dpurple}{HTML}{A020F0} \definecolor{dblue}{rgb}{0,0,1}
\definecolor{hlcolor}{rgb}{1,1,0.8}

\creflabelformat{equation}{#2#1#3}

\def\headline#1{{\hrulefill\quad\lower.3em\hbox{#1}\quad\hrulefill}}

 \sethlcolor{hlcolor}

\def\vectwo#1#2{\left(\begin{array}{c}  #1 \\ #2 \end{array}\right)}

\def\mat#1#2#3#4{\left(\begin{array}{cc} #1 & #2 \\ #3 & #4 \end{array}\right)}

\def\msub#1{{\mathop{\rm #1}}}

\def\lla{\left\langle}
\def\rra{\right\rangle}

\usepackage{lipsum}
\newenvironment{Figure}
  {\par\medskip\noindent\minipage{\linewidth}}
  {\endminipage\par\medskip}

\renewcommand\b\mathbf

\newcommand{\be}{\begin{equation}}
\newcommand{\ee}{\end{equation}}
\newcommand{\bea}{\begin{eqnarray}}
\newcommand{\eea}{\end{eqnarray}}
\newcommand{\bs}{\begin{split}}
\newcommand{\bes}{\begin{equation}\begin{split}}
\newcommand{\ees}{\end{split} \end{equation}}
\newcommand{\es}{\end{split}}

{}

\newcommand{\req}[1]{Eq.~(\ref{#1})}

\newcommand{\vc}[1]{{\mathbf{#1}}}
\newcommand{\vct}[1]{{\boldsymbol{#1}}}

\newcommand{\vr}{\vct{r}}

\newcommand{\vmu}{\vct{u}}

\newcommand{\vy}{\vc{y}}

\newcommand{\vv}{\vc{v}}
\newcommand{\vu}{\vc{u}}

\newcommand{\vg}{\vc{g}}
\newcommand{\vJ}{\vc{J}}
\newcommand{\vW}{\vc{W}}
\newcommand{\vI}{\vc{I}}
\newcommand{\vtW}{\vc{\widetilde W}}
\newcommand{\vtI}{\vc{\widetilde I}}
\newcommand{\wee}{W_{\msub{EE}}}
\newcommand{\wei}{W_{\msub{EI}}}
\newcommand{\wie}{W_{\msub{IE}}}
\newcommand{\wii}{W_{\msub{II}}}

\renewcommand{\a}{A}
\renewcommand{\b}{B}
\newcommand{\jab}{J_{\msub{AB}}}
\newcommand{\kab}{K_{\msub{AB}}}
\newcommand{\tauab}{\tau_{\msub{AB}}}

\newcommand{\wab}{W_{\msub{AB}}}
\newcommand{\wex}{W_{\msub{EX}}}
\newcommand{\wix}{W_{\msub{IX}}}
\newcommand{\re}{r_{\msub{E}}}
\newcommand{\gge}{g_{\msub{E}}}
\newcommand{\ri}{r_{\msub{I}}}

\newcommand{\rss}{\vr_{\msub{SS}}}
\newcommand{\yss}{\vy_{\msub{SS}}}

\newcommand{\rb}{r_{\msub{B}}}
\newcommand{\rx}{r_{\msub{X}}}
\newcommand{\ry}{r_{\msub{Y}}}
\newcommand{\taue}{\tau_{\msub{E}}}
\newcommand{\tauy}{\tau_{\msub{y}}}
\newcommand{\ke}{K_{\msub{E}}}
\newcommand{\ki}{K_{\msub{I}}}

\newcommand{\ky}{K_{\msub{Y}}}
\newcommand{\je}{J_{\msub{E}}}
\newcommand{\jy}{J_{\msub{Y}}}
\newcommand{\nE}{n_{\msub{E}}} 
\newcommand{\nI}{n_{\msub{I}}}

\newcommand{\ny}{n_{\msub{Y}}}
\newcommand{\nb}{n_{\msub{B}}}
\newcommand{\iE}{I_{\msub{E}}} 
\newcommand{\iI}{I_{\msub{I}}}

\newcommand{\mue}{\mu_{\msub{E}}}
\newcommand{\mui}{\mu_{\msub{I}}}
\newcommand{\muepost}{u_{\msub{E}}}
\newcommand{\muipost}{u_{\msub{I}}}

\newcommand{\mux}{\mu_{\msub{X}}}
\newcommand{\muy}{\mu_{\msub{Y}}}

\newcommand{\sigmae}{\sigma_{\msub{E}}}

\newcommand{\sigmay}{\sigma_{\msub{Y}}}
\newcommand{\sigmax}{\sigma_{\msub{X}}}
\newcommand{\sigmai}{\sigma_{\msub{I}}}
\newcommand{\oneoversqrtk}{\frac{1}{\sqrt{K}}}
\newcommand{\Kn}{K}

\newcommand{\om}{\mathcal{O}}
\newcommand{\pow}{p}
\def\mysec#1{{\noindent \bf #1}}

\def\eg{{\it e.g.}}
\def\ie{{\it i.e.}}

\title{What is the dynamical regime of cerebral cortex?}

\author[1,2]{Yashar Ahmadian}

\author[3]{Kenneth D.\ Miller}

\affil[1]{Institute of Neuroscience, Departments of Biology and Mathematics,
\mbox{University of Oregon, OR, USA}} 

\affil[2]{As of March 2020: Computational and Biological Learning Lab, Department of Engineering, University of Cambridge, Cambridge, UK}

\affil[3]{Center for Theoretical Neuroscience, Swartz Program in Theoretical Neuroscience, Kavli Institute for Brain Science, and Dept. of Neuroscience, College of Physicians and
Surgeons and Morton B.\ Zuckerman Mind Brain Behavior Institute, \mbox{Columbia University, NY, USA}} 

\begin{document}

\maketitle

\begin{abstract} \normalsize
\noindent Many studies have shown that the excitation and inhibition received by cortical neurons remain roughly balanced across many conditions. A key question for understanding the dynamical regime of cortex is the nature of this balancing. Theorists have shown that network dynamics can yield systematic cancellation of most of a neuron's excitatory input by inhibition.  We review a wide range of evidence pointing to this cancellation occurring in a regime in which the balance is loose, meaning that the net input remaining after cancellation of excitation and inhibition is comparable in size to the factors that cancel, rather than tight, meaning that the net input is very small relative to the cancelling factors. This choice of regime has important implications for cortical functional responses, as we describe: loose balance, but not tight balance, can yield many nonlinear population behaviors seen in sensory cortical neurons,  allow the presence of correlated variability, and yield decrease of that variability with increasing external stimulus drive  as observed across multiple cortical areas.
\phantomsection\addcontentsline{toc}{section}{Abstract}\noindent
\end{abstract}

\begin{multicols}{2}

\phantomsection \addcontentsline{toc}{section}{Introduction}

In what regime does cerebral cortex operate? This is a fundamental
question for understanding cerebral cortical function. The concept of
a ``regime'' can be defined in various ways. Here we will focus on a
definition in terms of the balance of excitation and inhibition: how
strong are the excitation and inhibition that cortical cells receive,
and how tight is the balance between them? As we will see, the answers
to these questions have important implications for the dynamical function of
cortex. 

We first consider several more fundamental distinctions in cortical regime. First, neurons may fire in a regular or irregular fashion, where regular firing refers to emitting
spikes in a more clock-like manner, while irregular firing refers to
emitting spikes in a more random manner, like a Poisson process. Cortex
appears to be in an irregular
regime \citep{Softky_Koch93,Shadlen_Newsome98}, though some areas are
less irregular than others
\citep{Maimon_Assad09}. Second, neurons may fire in a synchronous
regime, meaning with strong correlations between the firing of
different neurons, or an asynchronous regime, meaning with weak (or
no) correlations. Cortical
firing, particularly in the
awake state, generally appears to be in an asynchronous regime
\citep{Cohen_Kohn11,Ecker_etal10,Ecker_etal14,Doiron_etal16}, although some conditions may show
more synchronous
firing \citep{Stevens_Zador98,DeWeese_Zador06,Poulet_Petersen08,Tan_etal14}. Thus, we will take cortex to be in an asynchronous irregular regime. \citet{Brunel00b} first defined conditions on networks of excitatory and inhibitory neurons that led them to operate in the asynchronous irregular regime.

A third distinction is whether a network goes to a stable fixed rate of firing for a fixed input (with noisy fluctuations about that fixed rate given noisy inputs), or whether it shows more complex behaviors, such as movement between multiple fixed points, oscillations, or chaotic wandering of firing rates. We will focus on the case of a single fixed point, which seems likely to reasonably approximate at least awake sensory cortex \citep[see discussion in][]{Miller16}. The fixed point is taken to be {\em stable}, meaning that the network dynamics cause firing rates to return to the fixed point levels after small perturbations.
Finally, for a given fixed point,  recurrent excitation may be strong enough to destabilize the fixed point in the absence of feedback inhibition; that is, if inhibitory firing were held frozen at its fixed point level, a small perturbation of excitatory firing rates would cause them to either grow very large or collapse to zero. In that case, the fixed point is stabilized by feedback inhibition, and the network is known as an inhibition-stabilized network (ISN). Alternatively, the recurrent excitation may be weak enough to remain stable even without feedback inhibition.  A number of studies have found strong evidence that at least primary visual and auditory cortices are ISNs both at spontaneous \citep{Sanzeni_etal19} and stimulus-driven \citep{Ozeki_etal09,Kato_etal17,Adesnik17} levels of activity.

Note that for some of the distinctions we describe between regimes, there is a sharp transition from one regime to the other, while for others the transition is gradual. We use the word ``regime" in either case to describe qualitatively different network behaviors.

The assumption that cortex is in an irregularly-firing regime (as well as its operation as an ISN)
strongly points to the need for some kind of balance between
excitation and inhibition. Stochasticity of cellular and
synaptic mechanisms \citep{Mainen_Sejnowski95,Schneidman_etal98,Odonnell_vanRossum14} and input correlations \citep{Stevens_Zador98,DeWeese_Zador06} may contribute to irregular firing. However, a number of authors have argued that, assuming inputs are un- or weakly-correlated, then irregular
firing will arise if the mean input to cortical cells is sub- or peri-threshold,
so that firing is induced by fluctuations from the mean rather than by
the mean itself
\citep{Tsodyks_Sejnowski95,vanVreeswijk_Sompolinsky96,Troyer_Miller97,Amit_Brunel97b}.  This is referred to as the fluctuation-driven regime, as opposed to the
mean-driven regime in which the mean input is strongly suprathreshold and
spiking is largely driven by integration of this mean input.  The
fluctuation-driven regime yields random, Poisson-process-like firing,
because fluctuations are equally likely to occur at any time, whereas the mean-driven regime yields regular firing.

Given the  strength of inputs to cortex (to be discussed below),
the mean excitation received by a strongly-responding cell is likely to be sufficient to
drive the cell near or above threshold. Therefore, for the mean input to
be subthreshold, the mean inhibition is likely to  cancel a significant portion
of the mean excitation; that is, the excitation and inhibition
received by a cortical cell should be at least roughly
balanced \citep{Tsodyks_Sejnowski95,vanVreeswijk_Sompolinsky96}. Consistent with
the idea that inhibition balances excitation, many experimental
investigations have suggested that cortical or hippocampal excitation and inhibition
remain balanced or inhibition-dominated across varying activity levels
\citep{Galarreta_Hestrin98,Anderson_etal00,Shu_etal03,Wehr_Zador03,Marino_etal05,Wu_etal06,Haider_etal06,Higley_Contreras06,Wu_etal08,Okun_Lampl08,Atallah_Scanziani09,Yizhar_etal11,Graupner_Reyes13,Haider_etal13,Zhou_etal14,Barral_Reyes16,Dehghani_etal16}.

Excitation and inhibition may be balanced in at least two ways. First, inhibitory and excitatory synaptic weights may be co-tuned, so that cells that receive more (or less) excitatory weight receive correspondingly more (or less) inhibitory weight \citep{Xue_etal14,Bhatia_etal19}. This does not ensure balancing of excitation and inhibition received across varying patterns of activity. Second, given sufficiently strong feedback inhibitory weights, the network dynamics may ensure that the mean inhibition and  mean excitation received by neurons remain balanced across patterns of activity, without requiring tuning of synaptic weights. Here, we will focus on this second, dynamic form of balancing. 

As we will discuss, theorists have described mechanisms by which
inhibition and excitation dynamically remain balanced, keeping neurons in the
fluctuation-driven regime, without any need for
fine tuning of parameters such as synaptic weights. This dynamical balance can be a ``tight balance'', which we define to 
mean that the excitation and inhibition that cancel are much larger
than the residual input that remains after cancellation, or a ``loose
balance'', meaning that the canceling inputs are 
 comparable in size to the remaining residual input (terms to describe balanced networks are not yet standardized; see Appendix 1 for comparison of our usage to other nomenclatures). The question of whether the
balance is tight or loose has important implications for the behavior
of the network. Here we will review these issues and argue that the
evidence is most consistent with a loosely balanced regime. 

\subsection*{A Theoretical Problem: How to Achieve Input Mean and Fluctuations That Are Both Comparable in Size to Threshold?}

How do cortical neurons stay in the irregularly firing
regime? There are two requirements to be in the fluctuation-driven regime, which yields irregular firing: (1) The mean input the neurons receive must be sub- or
peri-threshold; (2) Input fluctuations must be sufficiently large to bring
neuronal voltages to spiking threshold sufficiently often to create reasonable
firing rates. We will measure the voltage effects of a neuron's inputs
in units of the voltage distance from the neuron's rest to threshold;
this distance, typically around 20 mV for a cortical cell \citep[\eg,][Fig.~3K]{Constantinople_Bruno13}, is equal to
1 in these units. Thus, a necessary condition for being in the irregularly firing
regime is that the voltage mean driven by the mean input (henceforth
abbreviated to ``the mean input'') should  have order of magnitude
1, which we write as $\om(1)$, or smaller. The second requirement above then dictates that the
voltage fluctuations driven by the input fluctuations from the mean  (henceforth abbreviated to ``input
fluctuations'') should also be $\om(1)$. In particular, this means that the
ratio of the mean input to the input fluctuations should  be $\om(1)$. (Note, we use the $\om()$ notation simply to indicate order of magnitude, and not in its more technical, asymptotic sense of the scaling with some parameter as that parameter goes to zero or infinity.)

Several authors have considered the requirements for these conditions
to be
true \citep{Tsodyks_Sejnowski95,vanVreeswijk_Sompolinsky96,vanVreeswijk_Sompolinsky98,Renart_etal10}.
Following these authors, we assume the network is composed of excitatory ($E$)
and inhibitory ($I$) neuron populations, which receive excitatory inputs from an
external ($X$) population. The latter could represent any cortical or
subcortical neurons outside the local cortical network, for example,
the thalamic input to an area of primary sensory cortex. As a simplified toy model of 
the assumption that the network is in the asynchronous irregular regime, we assume
that the cortical cells fire as Poisson processes without any
correlations between them, as do the cells in the external population.

Suppose that a neuron receives $\ke$ excitatory inputs. Suppose these inputs produce EPSPs that have an exponential time course, with mean amplitude $\je$ and time constant $\taue$,  and have mean rate $\re$. Then the mean depolarization produced by these excitatory inputs is $\je\ke\re\taue$. Defining $\nE=\re\taue$ to be the mean number of spikes of an input in time $\taue$, we find that the mean excitatory input to the neuron is
\be\label{muEeq}
\mue = \je \ke \nE 
\ee
Let $\sigmae$ denote the standard deviation (SD) of fluctuations in
this input. Assuming the spike counts of
pre-synaptic neurons are uncorrelated, their spike count
variances just add. Because they are firing as Poisson
processes, the variance in a neuron's spike count in time $\taue$ is equal
to its mean spike count $\nE$. Thus, the variance of input from one pre-synaptic
neuron is $\je^2 \nE$, and so the variance in the total input is
$\ke \je^2 \nE$ and
\be\label{sigmaEeq}
\sigmae = \je\sqrt{\ke \nE}
\ee
Therefore the ratio of the mean to the SD of the neuron's excitatory
input is
\be \label{ratioeq}
\frac{\mue}{\sigmae} = \sqrt{\ke \nE},
\ee
independent of $\je$.
Similar reasoning about the neuron's inhibitory or external input leads to all the same expressions, except with $E$ subscripts replaced with $I$ or $X$ subscripts to represent quantities describing the inhibitory or external input the cell receives.

Again assuming that the different populations are uncorrelated so that
their contributed variances add, 
the {\em total} or {\em net} input the neuron receives  has mean, $\mu$, and standard deviation, $\sigma$, given by: 
\begin{align}
  \mu &= \mue + \mux - \mui \label{muaeq}\\
  \sigma &= \sqrt{\sigmae^2 + \sigmax^2 + \sigmai^2} \label{sigmaaeq}
\end{align}
We imagine that $\ke$ and $\ki$
are the same order of magnitude, $\om(K)$ for some number
$K$, and similarly $\nE$ and $\nI$ are $\om(n)$ for some number $n$.  We also assume $\mux$ and $\sigmax$ are the same order of magnitude as $\mue or \mui$ and $\sigmae or \sigmai$, respectively, or smaller.
Then, if $\sqrt{Kn}$ is $\om(1)$, 
both $\mu$ and $\sigma$ can simultaneously be made $\om(1)$ with
suitable choice of the $J$'s 
(generically, \ie\ barring special cases in which the elements of
$\mu$ precisely cancel). 
This means that
the irregularly-firing regime is self-consistent: having assumed that neurons are in
the irregularly-firing regime, we arrive at expressions for the mean and SD of their input
that indeed can keep the network in this regime. 

If $K$ is very large, however -- large enough that $\sqrt{Kn}\gg 1$ -- then the
ratio of the mean to the SD of each type of input, and hence of the
net input, is much greater
than 1. Van Vreeswijk and Sompolinsky (1996,1998) considered the case of such very
large $\Kn$ and showed how the network could remain in the AI
regime.  They proposed choosing the $J$'s proportional to
$\oneoversqrtk$, so that the standard deviations $\sigmae$ and $\sigmai$ are $\om(1)$
(Eq.~\ref{sigmaEeq}), but then by Eq.~\ref{ratioeq} the means $\mue$ and $\mui$ are large, $\om(\sqrt{\Kn})$. 
Then, for the  neurons to be in the asynchronous irregular regime,
inhibitory input, $-\mui$, must cancel or ``balance'' a sufficient portion of the
excitatory input, $\mue+\mux$, so that the mean input, $\mu$, is $\om(1)$. 

If a neuron's mean excitatory and  inhibitory inputs very precisely cancel each
other, so that the mean net input $\mu$ is much smaller than either of
these factors alone, we say there is {\em tight balance}. If the net
input is more comparable in size to the factors that are cancelling, we
call this {\em loose balance}. The two cases may be distinguished by
the size of a dimensionless {\em balance index} $\beta$:
\be
\beta = \frac{\left|\mu\right|}{\mue+\mux}
\label{bi}
\ee
Note that, using the above analysis, in the limit of large $K$ considered by  Van Vreeswijk and Sompolinsky (1996,1998), $\beta\sim \oneoversqrtk$.
Tight balance means that the balance index is very small, $\beta \ll
1;$ 
in loose balance, the index is not so small, say $0.1<\beta<1$, very roughly. 
As we will see, whether the network is in tight or loose
balance has important implications for the network's behavior and
computational ability. 
(Note that, in general, the degree of balance can be different in different neurons in the same network. In the above discussion we assumed that different neurons of the same $E/I$ type are statistically equivalent, \eg, in terms of the number and activity of presynaptic inputs; this is the case in the randomly connected network of  Van Vreeswijk and Sompolinsky (1996,1998). In that case $\beta$ would not vary systematically between neurons of the same type.)

\subsection*{The Tightly Balanced Solution}

Van Vreeswijk and Sompolinsky (1996,1998) showed that, for very large $K$, and hence very large 
$Kn$, and all $J$'s $\propto \oneoversqrtk$, the network dynamics would produce a tightly balanced solution provided only that some mild (inequality) conditions on the
weights are satisfied, without any requirements for fine tuning.  This
is known more generally as the ``balanced network'' solution. To
understand this solution, we define the mean number of inputs, PSP amplitude, and time constant from population $\b$ ($\b\in\{E,I,X\}$) to a neuron in population $\a$ ($\a\in\{E,I\}$) to be $\kab$, $\jab$ and $\tauab$ respectively. We define the mean effective 
weight from population $\b$ to a neuron in population $\a$ as $\wab=\jab
\kab \tauab$. Letting $\rb$ be the average firing rate of population $B$, then $\wab\rb=\jab\kab\nb$, the mean input from population $B$ to population $A$. We assume that $\wab\rb=\om(\sqrt{\Kn})$ for all $\a,\b$.
The requirements for balance are then that the mean net input to both
excitatory and inhibitory cells, $\muepost$ and $\muipost$ respectively, are
$\om(1)$, where (from Eqs.~\ref{muEeq},\ref{muaeq}),
\begin{align}
  \muepost &= \wee \re - \wei \ri+\wex \rx  \label{muEeq2}\\
  \muipost &= \wie \re - \wii \ri+ \wix \rx  \label{muIeq2}
\end{align}
If we define the external inputs to the network $\iE=\wex \rx$,
$\iI=\wix \rx$, then these equations can be written as the vector
equation 
\be
\vmu=\vW\vr+\vI \label{veceq}
\ee
where $\vmu\equiv\vectwo{\muepost}{\muipost}$,
$\vr\equiv\vectwo{\re}{\ri}$, $\vI\equiv\vectwo{\iE}{\iI}$, and $\vW$
is the weight matrix $\vW=\mat{\wee}{-\wei}{\wie}{-\wii}$.

The balanced network solution arises by noting that the left side of
Eq.~\ref{veceq} is very small ($\om(1)$) relative to the individual
terms on the right ($\om(\sqrt{\Kn})$). So we first find an approximate
solution $\vr_0$ to Eq.~\ref{veceq} in which the small left side is
replaced by $0$ to yield the equation for perfect balance, \ie\ all
inputs perfectly cancelling: $\vW\vr_0+\vI=0$, or
$\vr_0=-\vW^{-1}\vI$, where $\vW^{-1}$ is the matrix inverse of
$\vW$. Note that $\vr_0$ is $\om(1)$, because the elements of $\vW$ and
$\vI$ are all the same order of magnitude, so their ratio is generically 
$\om(1)$. We can then write $\vr$ as an expansion in powers of
$\oneoversqrtk$, $\vr=\vr_0+\frac{\vr_1}{\sqrt{\Kn}}+\ldots$, where
$\vr_0$, $\vr_1$, $\ldots$ are all $\om(1)$, to obtain a consistent
solution: $\vmu=\frac{\vW \vr_1}{\sqrt{\Kn}} + \ldots$ where the first
term on the right is $\om(1)$, as desired, and the remaining terms are
very small ($\om\left(\oneoversqrtk\right)$ or smaller). The authors showed that,
with some mild general conditions on the weights $\vW$ and inputs
$\vI$, this tightly balanced solution would be the unique stable
solution of the network dynamics. That is, for a given fixed input
$\vI$, the network's excitatory/inhibitory dynamics will lead it to
flow to this balanced solution for the mean rates: $\vr=-\vW^{-1}\vI +
\om\left(\oneoversqrtk\right)$. 

We immediately see two points about the tightly balanced ($\sqrt{Kn}\gg
1$) solution:
\begin{enumerate}
  \item {\em Mean population responses are linear in the inputs}. $-\vW^{-1}\vI$ is a
    linear function of the input $\vI$. Tight balance implies
    that nonlinear corrections to $\vr\approx \vr_0=-\vW^{-1}\vI$ are very small relative
    to this linear term, except for very small external inputs, so mean response $\vr$ is for practical purposes a
    linear function of the input.

  \item {\em External input must be large relative to the net
    input and to the distance from rest to threshold.} The external input $\vI$ must have the same order of magnitude
    as the recurrent input $\vW\vr_0$, so that balance can occur,
    $\vW\vr_0=-\vI$, with rates that are $\om(1)$. If $\vI$ were
    smaller, the firing rates $\vr\approx \vr_0$ would correspondingly
    be unrealistically small. 
\end{enumerate}

In the above treatment
we focused on population-averaged responses, $\re$ and $\ri$.
We emphasize that the balancing only applies to the mean input across neurons of each type, and leaves unaffected input components with mean of zero across a given type; while the mean input is very large in the tightly balanced regime, zero-mean input components can be $\om(1)$ and yet elicit $\om(1)$ responses in individual neurons  (see \eg\ \cite{Hansel_vanVreeswijk12, Pehlevan_Sompolinsky14, 
Sadeh_Rotter15}). 
Furthermore, even in
the tightly balanced regime, individual neurons can exhibit
nonlinearities in their responses, but these are washed out at the
level of population-averaged responses. 
We also note that synaptic nonlinearities, \eg\ synaptic depression, which
were neglected here, can allow a tightly balanced state with nonlinear population-averaged
responses \citep{Mongillo_etal12}. 

\subsection*{A Loosely Balanced Regime}
As we have seen, if $Kn$ is $\om(1)$, the mean and the fluctuations of the input that neurons receive can both be $\om(1)$ without requiring any balancing. Given that there is both excitatory and inhibitory input, there will always be some input cancellation or "balancing" -- some portion of the input excitation will be cancelled by input inhibition, leaving some smaller net input. When $Kn$ is $\om(1)$, all of these quantities -- the excitatory input, the inhibitory input, and the net input after cancellation -- will generically be $\om(1)$, and thus balancing is ``loose" --  the factors that cancel and the net input after cancellation are of comparable size, and the balance index $\beta$ is not small.

However, the fact that there is some inhibition that cancels some excitation does not by itself imply interesting consequences for network behavior. We will use the term "loosely balanced solution" to refer more specifically to a solution having two features: (1) the dynamics yields a systematic cancellation of excitation by inhibition like that in the tightly balanced solution. In particular, in the loosely balanced networks on which we will focus, a signature of this systematic cancellation is that  the net input a neuron receives grows sublinearly as a function of its external input (we will make this more precise below);  (2) this cancellation is ``loose", as just described. As we will discuss, such a loosely balanced solution produces various specific nonlinear network behaviors that are  observed in cortex. 

Ahmadian et al. (2013) showed that such a loosely balanced solution would
naturally arise from E/I dynamics for recurrent weights and external
inputs that are not large, provided
that the neuronal input/output function, determining firing rate vs.\ input level, is
supralinear (having ever-increasing slope) over the neuron's dynamic
range. They modeled this supralinear input/output function as a power law with power greater than 1 (Fig.~\ref{powerlawfig}). Such a power-law input-output function is theoretically expected for a spiking neuron when firing is induced by input fluctuations rather than the input mean \citep{MIller_Troyer02,Hansel_vanVreeswijk02}, and is observed in intracellular recordings over the full dynamic range of neurons in primary visual cortex (V1) \citep{Priebe_Ferster08}. Of course, a neuron's input/output function must ultimately saturate but, at least in V1, the neurons do not reach the saturating portion of their input/output function under normal operation. For the loosely balanced solution to arise, some general conditions on the weight matrix, similar to those for the tightly balanced network solution but less restrictive, must also be satisfied.

\begin{Figure}\hspace{0cm}
\begin{center}
\includegraphics[width=\linewidth,angle=0]{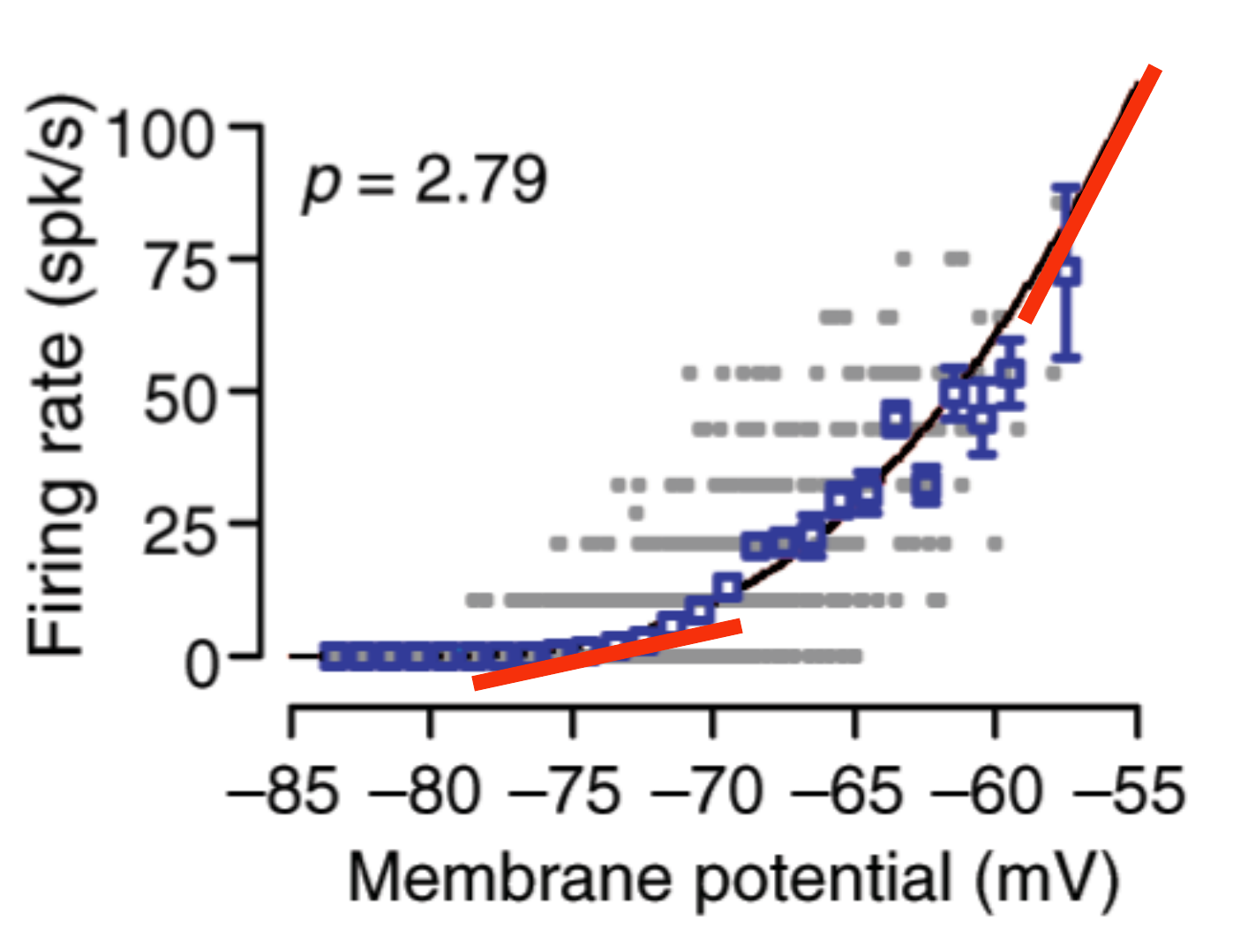}
\captionof{figure}{{\bf The supralinear (power-law) neuronal  transfer function.} The transfer function of neurons in cat V1 is non-saturating in the natural dynamic range of their inputs and outputs, and is well fit by  a supralinear rectified power-law with exponents empirically found to be in the range 2-5. Such a curve exhibits increasing input-output gain (i.e. slope, indicated by red lines) with growing inputs, or equivalently with increasing output firing rates. Gray points indicate a studied neuron's average membrane potential and firing rate in 30ms bins; blue points are averages for different voltage bins; and black line is fit of power law, $r=[V-\theta]_+^p$, where $r$ is firing rate, $V$ is voltage, $[x]_+=x$, $x>0$, =0 otherwise; $\theta$ is a fitted threshold; and $p$, the fitted exponent, here is 2.79. Figure modified from \citep{Priebe_etal04}.  }
\label{powerlawfig}
\end{center}
\end{Figure} 	

In the presence of a supralinear input/output function, the loosely balanced solution arises as follows.  Whereas previously we considered the effects of increasing $\Kn$ when recurrent and external inputs were all $\om(\sqrt{\Kn})$, now we consider the more biological case of increasing external input (\ie, stimulus) strength while recurrent weights are at some fixed level.
The supralinear input/output function means that a neuron's gain -- its change in output for a given change in input -- is continuously increasing with its activation level. This in turn means that effective synaptic strengths are increasing with increasing network activation. The effective synaptic strength measures the change in the postsynaptic cell's firing rate per change in presynaptic firing rate. This is the product of the actual synaptic strength -- the change in postsynaptic input induced by a change in presynaptic firing -- and the postsynaptic neuron's gain, hence the effective synaptic strengths increase with increasing gains.

The increasing effective synaptic strengths lead to two regimes of network operation. For very weak external drive and thus weak network activation, all effective synaptic strengths are very weak, for both externally-driven and network-driven synapses. External drive to a neuron is delivered monosynaptically, via the weak externally-driven synapses. In contrast, assuming that the network is inactive in the absence of external input, network drive involves a chain of two or more weak synapses: the weak externally driven synapses activate cortical cells, which then drive the weak network-driven synapses. From the same principle that $x^2\ll x$ when $x\ll 1$, the network drive is therefore much weaker than the external drive. Thus, the input to neurons is dominated by the external input, with only relatively small contributions from recurrent network input. In sum, in this weakly-activated regime, the neurons are {\em weakly coupled}, largely responding directly to their external input with little modification by the local network. 

With increasing external (stimulus) drive and thus increasing network activation, the gains and thus the effective synaptic strengths grow. This causes the relative contribution of network drive to grow until the network drive is the dominant input. At some point, the effective $E\rightarrow E$ connections become strong enough that the network would be prone to instability -- a small upward fluctuation of excitatory activities would recruit sufficient recurrent excitation to drive excitatory rates still higher, which if unchecked would lead to runaway, epileptic activity (and to ever-growing effective synaptic strengths and thus ever-more-powerful instability). However, if feedback inhibition is strong and fast enough, the inhibition will stabilize the network, that is, it becomes an ISN. This stabilization is achieved by a loose balancing of excitation and inhibition, as we will explain in more detail below. Thus, in this more strongly-activated regime, the neurons are {\em strongly coupled} and are {\em loosely balanced}. Note that the input driving spontaneous activity may already be strong enough to obscure the weakly coupled regime, as suggested by the finding that V1 under spontaneous activity is already an ISN \citep{Sanzeni_etal19}. As in the tightly balanced network, the network's excitatory/inhibitory dynamics lead it to flow to this loosely balanced solution, without any need for fine tuning of parameters. Because this mechanism involves stabilization, by inhibition, of the instability induced by the supralinear

\end{multicols}
\begin{figure}[!t]\hspace{0cm}
\begin{center}
\includegraphics[angle=0]{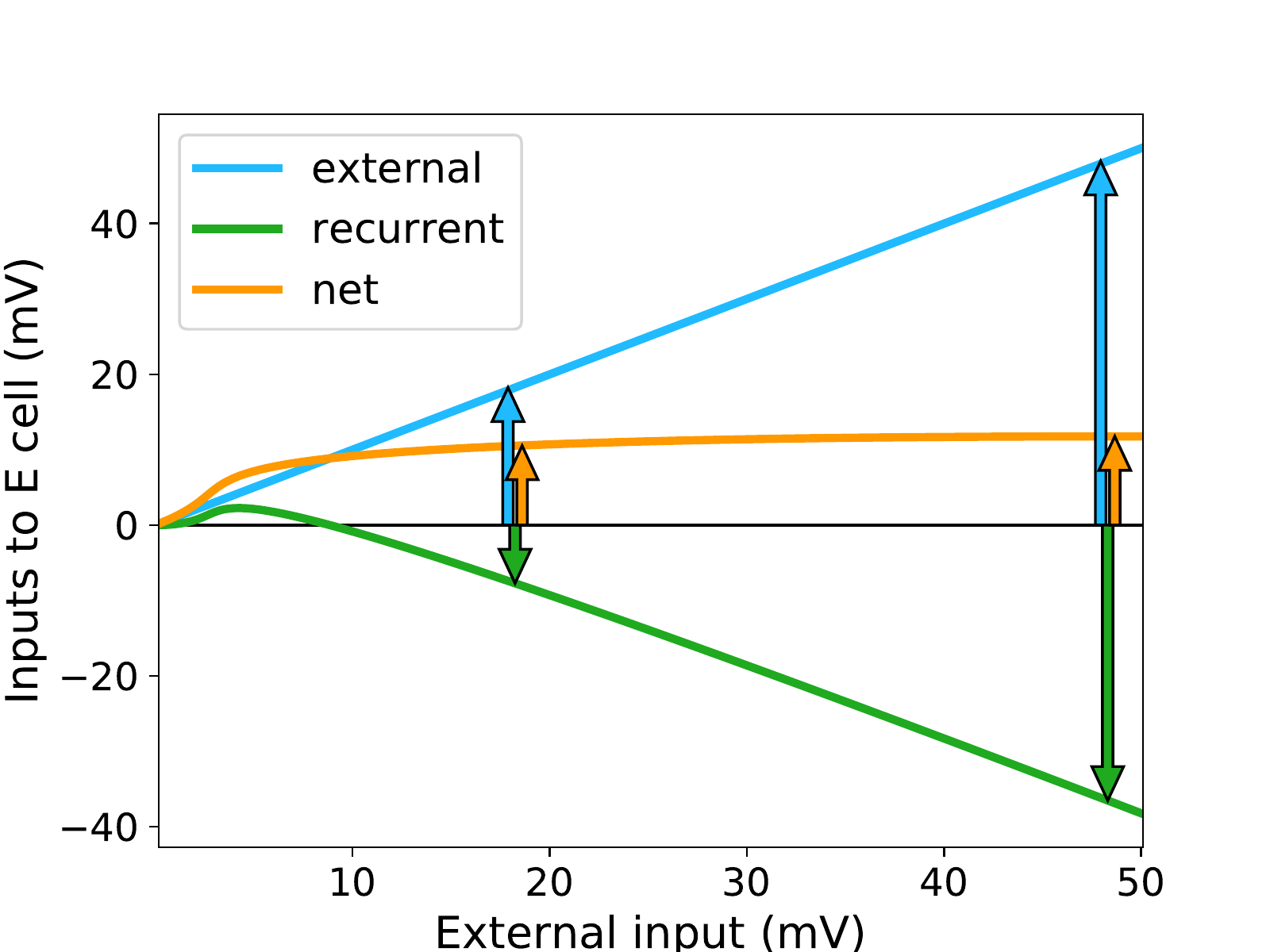}
\caption{{\bf Loose vs. tight balance.} We illustrate the external (blue), recurrent or within-network (green) and net (orange, equal to external plus recurrent) inputs to a typical excitatory cell in a Stabilized Supralinear Network. At all biological ranges of external input (stimulus) strength, the balance is loose, as exhibited by the left set of three arrows (representing the external, recurrent and net inputs): the net input is comparable in size to the other two.  Nevertheless the balance systematically tightens with increasing external input (right set of arrows), as the net input grows only sublinearly with growing external input strength. At very high (possibly non-biological) levels of external input, the balance can become very tight, with the net input much smaller in magnitude than the external and recurrent inputs.}
\label{fig-2}
\end{center}
\end{figure} 	
\begin{multicols}{2}

input/output function of individual neurons along with $E\rightarrow E$ connections, it has been termed the {\em Stabilized Supralinear Network} (SSN) \citep{Ahmadian_etal13,Rubin_etal15}. 

To describe the mathematics of this mechanism, we again consider an excitatory and an inhibitory population along with external input. We define the vectors $\vr$, $\vu$ and $\vI$ and the matrix $\vW$ as before. Then the power-law input/output function means that the network's steady state firing rate 
$\rss$ for a steady input $\vI$ satisfies
\be\label{ssnSS}
\rss= k (\vmu)_+^\pow = k(\vW\rss+\vI)^\pow_+
\ee
where $(\vv)_+$ is the vector $\vv$ with negative elements set to zero, $(\vv)_+^\pow$ means raising each element of $(\vv)_+$ to the power $\pow$, $\pow$ is a number greater than 1 (typically, 2 to 5, \citealp{Priebe_Ferster08}), 
and $k$ is a constant with units $\frac{Hz}{(mV)^\pow}$ (and the units of $\vW$, $\vr$, and $\vI$ are $\frac{mV}{Hz}$, $Hz$, and $mV$ respectively). 
It is convenient to absorb $k$ into effective weights and inputs by writing $\vtW=k^{1/\pow}\vW$, $\vtI=k^{1/\pow}\vI$, so the equation becomes
\be
\rss=(\vtW\rss+\vtI)^\pow_+
\ee
If we let $\psi=\|\vtW\|$ represent a norm of $\vtW$ (think of it as a typical total E or I recurrent weight received by a neuron), and similarly let $c= \|\vtI\|$ represent a typical input strength, then it turns out the network regime is controlled by the dimensionless parameter
\be
\alpha=\psi c^{\pow-1}.
\ee
As we increase the strength of external drive and thus of network activation, $c$ and thus $\alpha$ is increasing. When $\alpha\ll 1$, the network is in the weakly coupled regime; for $\alpha\gg 1$, the network is in the strongly coupled regime; and the transition between regimes generically occurs when $\alpha$ is $\om(1)$ \citep{Ahmadian_etal13}. 

The loosely-balanced solution then turns out to be of the form 
\be
\vr=-\vW^{-1}\vI+\frac{c}{\psi}\left(\frac{\vr_1}{\alpha^{1/\pow}}+\ldots \right)
\label{loose}
\ee
where $\vr_1$ is dimensionless and $\om(1)$, and the higher-order terms (indicated by $\ldots$) involve higher powers of $\frac{1}{\alpha^{1/\pow}}$ (see Appendix 2). Equation \ref{loose} is precisely the same equation as for the tightly balanced solution, in the case that the input/output function is a power law. In the tightly balanced network, $\psi$ and $c$ are both $\om(\sqrt{\Kn})$, so $\alpha$ is $\om((\Kn)^{\pow/2})$, \ie\ very large, and the $\frac{1}{\alpha^{1/\pow}}$ in the second term becomes $\om\left(\oneoversqrtk\right)$, as expected. The loosely-balanced
solution arises, however, when $\alpha$ is $\om(1)$. In particular, in the biological case of fixed weights but increasing stimulus drive, and given 
the supralinear neuronal input/output functions, the same E/I dynamics that lead to the tightly balanced solution when inputs are very large will already yield a loosely balanced solution when inputs are $\om(1)$. The conditions for this loosely balanced solution to arise are further discussed in Appendix 2.

The fact that the solution is loosely balanced can be seen by computing the balance index, $\beta$ (Eq.~\ref{bi}). The network excitatory drive is $\om(\psi \vr)$, the external drive is $\om(c)$, and because the first term on the right side of Eq.~\ref{loose} cancels the external input, the net input after balancing is $\vW$ (which is $\om(\psi)$) times the 2nd term on the right side of Eq.~\ref{loose}, or $\om\left(\frac{c}{\alpha^{1/\pow}}\right)=\om\left(\left(\frac{c}{\psi}\right)^{1/\pow}\right)$.  Since $1/\pow < 1$, the net input thus grows sublinearly with growing external input strength, $c$, as illustrated in Fig.~\ref{fig-2}. Moreover, it follows that the balance index (Eq.~\ref{bi}) is 
$\om\left(\frac{c/\alpha^{1/\pow}}{c+\psi\vr}\right)$ which is $\om\left(\frac{1}{\alpha^{1/\pow}}\right)$ (assuming the order of magnitude of the recurrent input, $\psi\vr$, is the same as or smaller than that of the external input strength, $c$). Again, for the tightly balanced solution this is very small, $\om\left(\oneoversqrtk\right)$, but the loosely balanced solution arises when this is $\om(1)$.

In more complex models (involving many neurons with structured connectivity and stimulus selectivity), 
loosely-balanced solutions still arise when $\alpha$ is $\om(1)$. That is, even in such cases the full nonlinear steady state equations, \req{ssnSS}, can yield biologically plausible solutions, and when that happens  the net inputs to activated neurons  grow sublinearly with growing external input strength, and balance indices are $\om(1)$. The case of structured networks with stimulus selectivity is further discussed in Appendix 2.

We can now see that the loosely balanced regime differs from the tightly balanced in the two points summarized previously:
\begin{enumerate}
\item {\em In the loosely balanced regime, mean population responses are nonlinear in the inputs.} This is because, when balance is loose, the second term in Eq.~\ref{loose}, which is not linear in the input, cannot be neglected relative to the first, linear term. In particular, the nonlinear population behaviors observed in the loosely balanced regime with a supralinear input/output function closely match the specific nonlinear behaviors observed in sensory cortex \citep{Rubin_etal15}, as we will discuss below.
\item {\em In the loosely balanced regime, external input can be comparable to the net input and to the distance from rest to threshold.} 
\end{enumerate}

\subsection*{What Regime Do Experimental Measurements Suggest?}

As we have seen above, the same model can give a loosely balanced solution (Eq.~\ref{loose}) when $\alpha$ is $\om(1)$ (\eg, when $c$ and $\psi$ are both $\om(1))$, but give a tightly balanced solution when $\alpha$ is large (\eg, when $c$ and $\psi$ are both $\om(\sqrt{\Kn})$). Which of these regimes is supported by experimental measurements?

\subsubsection*{Measurements of Biological Parameters}

\mysec{How large is $\sqrt{Kn}$?}
We saw in Eq.~\ref{ratioeq} that the ratio of the mean to the standard deviation, $\muy/\sigmay$, of the input of type $Y$ ($Y\in\{E,I,X\}$) received by a neuron is equal to $\sqrt{\ky \ny}$, where $\ky$ is the number of inputs of type $Y$ a given neuron receives and $\ny$ is the average number of spikes one of these inputs fires in a PSP decay time $\tauy$

\end{multicols}
\begin{table}[h]
\centering
\begin{tabular}{|l|c|c|c|} 
\hline
&$\ke=200$ & $\ke=1000$ & $\ke=5000$ \\
\hline
$\re=0.1$ Hz &0.6 &1.4 &3.2 \\
\hline
$\re=1$ Hz &2.0 &4.5 &10.0 \\
\hline
$\re=10$ Hz &6.3 &14.1 &31.6 \\
\hline
\end{tabular}
\caption{Values of $\sqrt{\ke\nE}$ for varying $\ke$ and $\re$, for $\tau=20$ms.}
\label{spines}
\end{table}

\begin{multicols}{2}

($\ny=\ry\tauy$, where $\ry$ is the average firing rate of one of these inputs). Here we estimate $\sqrt{\ke\nE}$. 

Note that $\sqrt{\ky\ny}$ is actually an upper bound for the ratio $\muy/\sigmay$ for a given input type, because we have neglected a number of factors that would increase fluctuations for a given mean. These include (i) correlations among inputs which, even if weak, can significantly boost the input SD, $\sigmay$, without altering input mean; (ii) variance in the weights, $\jy$, which would increase the estimate of $\sigmay$ by a factor $\sqrt{\lla \jy^2\rra/\lla \jy\rra^2}$;  and (iii) network effects that can amplify input variances by creating firing rate fluctuations,
 although this amplification may be small for strong stimuli \citep{Hennequin_etal18}. Furthermore, the ratio $\mu/\sigma$ of {\em total} input is expected to be smaller than the ratio $\muy/\sigmay$ for any single type. This is because $\sigma^2$ involves the sum of three variances (Eq.~\ref{sigmaaeq}), while $\mu$ involves a difference of one mean from the sum of two others (Eq.~\ref{muaeq}), representing the effect of loose balancing.

Given these considerations, we are primarily concerned with estimating the overall magnitude of $\sqrt{\ke\nE}$ rather than detailed values. If this magnitude is very much larger than the observed $\mu/\sigma$ ratio {\em in vivo}, then tight balancing may be needed to explain the {\em in vivo} ratio. To estimate the {\em in vivo}  $\mu/\sigma$ ratio, we note that, in anesthetized cat V1, $\sigma$ varies from 1 to 7 $mV$ and $\mu$ ranges from 0 to 15 $mV$ (20 $mV$ in one case) for a strong stimulus \citep{Finn_etal07,Sadagopan_Ferster12}.  While these authors did not give the paired $\mu$ and $\sigma$ values for individual cells, it seems reasonable to guess from these values that the value of $\mu/\sigma$ for the total input to these cells is generally in the range 0$-$15.  \citet{Finn_etal07} also reported that, at the peak of a simple cell's voltage modulation to a high-contrast drifting grating, the ratio $\sigma/\mu$ had an average value of about 0.15 (here, we are taking $\mu$ to be the mean voltage at the peak). This suggests that the average value of $\mu/\sigma$ at peak activation is around $1/0.15 = 6.7$.

How large is $\sqrt{\ke\nE}?$ In a study of input to excitatory cells in layer 4 of rat S1 \citep{Schoonover_etal14}, the EPSP decay time $\tau_{\msub{E}}$ was around 20 ms. From 1800 to 4000 non-thalamic-recipient spines were found on studied cells which, with an estimated average of 3.4 synapses per connection between layer 4 cells \citep{Feldmeyer_etal99}, corresponds to a $\ke$ -- the number of other cortical cells providing input to one cell -- of 530 to 1200. If $\re$ is expressed in Hz, then $\sqrt{\ke\nE}$ ranges from $3.3\sqrt{\re}$ to $4.9\sqrt{\re}$. Thus, even if average input firing rates were 10 Hz, which would be very high for rodent S1 \citep{Barth_Poulet12} (note that the average is over all inputs, not just those that are well driven in a given situation, and 
so is likely far below the rate of a well-driven neuron), these ratios would be 10.4$-$15.5. For more realistic rates of 0.1$-$1Hz \citep{Barth_Poulet12}, these ratios would be 1.0$-$4.9. All of these are comparable in magnitude to observed {\em in vivo} levels of $\mu/\sigma$. 

More generally, estimates across species and cortical areas of the number of spines on excitatory cells, and thus of the number of excitatory synapses they receive, range from 700 to 16,000, with numbers increasing from primary sensory to higher sensory to frontal cortices \citep{Elston03,Elston_Manger14,Elston_Fujita14,Amatrudo_etal12}. Estimates of the mean number of synapses per connection between excitatory cells range from 3.4 to 5.5 across different areas and layers studied \citep{Fares_Stepanyants09,Feldmeyer_etal99,Feldmeyer_etal02,Markram_etal97b}. These numbers yield a $\ke$ of 130 to 4700. In Table~\ref{spines}, we show the value of $\sqrt{\ke\nE}$ for $\ke$ ranging from 200 to 5000 (rounded upward to bias results most in favor of a need for tight balancing) and for rates $\re$ of 0.1 to 10 Hz. The results are all comparable to the $\mu/\sigma$'s observed {\em in vivo}, except for the most extreme case considered ($\ke=5000$, $\re=10$Hz), and even that case is only off by a factor of 2. Thus, the numbers strongly argue that tight balancing is not needed for the ratio of voltage mean to variance to have values as observed {\em in vivo}.

\mysec{External input is comparable in strength to net input.} Several studies have silenced cortical firing while recording intracellularly to determine the strength of external input, with cortex silenced, relative to the net input with cortex intact. These find the external input to be comparable to the net input, consistent with the loose balance scenario, rather than much larger as the tight balance scenario requires.

\citet{Ferster_etal96} cooled V1 and surrounding V2 in anesthetized cats to block spiking of almost all cortical cells, both excitatory and inhibitory, leaving axonal transmission (e.g. of thalamic inputs) intact, though with weakened release. By measuring the size of EPSPs evoked by electrical stimulation of the thalamic lateral geniculate nucleus (LGN) in thalamic-recipient cells in layer 4 of V1, they could estimate the degree of voltage attenuation of EPSPs induced by cooling. Correcting for this attenuation, they estimated that the first harmonic voltage response to an optimal drifting luminance grating stimulus of a layer 4 V1 cell was, on average, about 1/3 as great with cortex cooled as with cortex intact, suggesting that the external input to cortex is {\em smaller} than the net input with cortex intact. \citet{Chung_Ferster98} and \citet{Finn_etal07} assayed the same question by using cortical shock to silence the local cortex for about 150 ms, during which time the voltage response to an optimal flashed grating was measured. They found that on average the transient voltage response in layer 4 cells with cortex silenced was about 1/2 the size of that with cortex intact \citep{Chung_Ferster98}, and more generally ranged from 0\% to 100\% of the intact cortical response \citep{Finn_etal07}. This again suggests that the external input to cortex is {\em smaller} than the net input, \ie\ the external input is $\om(1)$, consistent with loose but not tight balance.

\mysec{Total excitatory or inhibitory conductance is comparable to threshold.} The above results suggest that depolarization due to thalamus alone is less than that induced by the combination of thalamic and cortical excitation plus cortical inhibition, \ie\ after cortical "balancing" has occurred. One can also ask what proportion of the total excitation is provided by thalamus. This has been addressed in voltage-clamp recordings in anesthetized mice by silencing cortex through light activation of parvalbumin-expressing inhibitory cells expressing channelrhodopsin. In layer 4 cells of V1 \citep{Lien_Scanziani13,Li_etal13} and primary auditory cortex (A1) \citep{Li_etal13b}, mean stimulus-evoked excitatory conductance with cortex silenced was estimated to be 30-40\% of that with cortex intact.

This tells us that the external and cortical contributions to excitation are comparable. How large are they compared to the excitation needed to depolarize the cell from rest to threshold, which is typically a distance of about 20 mV \citep{Constantinople_Bruno13}? With cortical spiking intact, these authors \citep{Lien_Scanziani13,Li_etal13b} found mean stimulus-evoked peak excitatory currents ranging from 60 to 150pA for various stimuli. Even assuming a membrane resistance of 200 M$\Omega$, which seems on the high end for {\em in vivo} recordings (\citet{Li_etal13b} reported input resistances of $150-200$ M$\Omega$), these would induce depolarizations of 12 to 30 mV; that is, the total excitatory current is comparable to threshold, \ie\ it is $\om(1)$.

A similar result can be found from decomposition of excitatory and inhibitory conductances from current-clamp recordings at varying hyperpolarizing current levels. In cat V1 cells for an optimal visual stimulus, one finds that peak excitatory and inhibitory stimulus-induced conductances, $g_E$ and $g_I$ respectively, are typically $<10$nS and almost always $<20$nS, on top of stimulus-independent conductances ($g_L$, for leak conductance)
around 10nS \citep{Anderson_etal00,Ozeki_etal09}. A study of response to whisker stimulation in rat barrel cortex found excitatory and inhibitory conductances of $\leq 5$ns \citep{Lankarany_etal16}.
The depolarization that the stimulus-induced excitatory conductance would induce by itself is $\frac{g_E}{g_E+g_L}V_E$, where $V_E$ is the driving potential of excitatory conductance, about 50 mV at spike threshold of around $-50$ mV \citep[\eg][]{Wilent_Contreras05b}. Using the cat V1 numbers, this means that the depolarization driven by excitatory conductance is typically $<25$ mV and almost always $<33$ mV. Hyperpolarization driven by the inhibitory conductance alone would be 0.4 to 0.6 times these values, given inhibitory driving force of $-20$ to $-30$ mV at spike threshold. These values are all quite comparable to the distance from rest to threshold, $\sim 20$ mV, that is, they are $\om(1)$.

\mysec{How large is the expected mean excitatory input?} We have seen that the expected mean depolarization induced by recurrent excitation is $J_EK_En_E$, where $J_E$ is the mean EPSP amplitude. Based on the measurements of \citet{Lien_Scanziani13} and \citet{Li_etal13}, discussed above, total excitation may be about 1.5 times greater than recurrent excitation. $J_E$ can be difficult to estimate, because some of the $K_E$ anatomical synapses may be very weak and not sampled in physiology, and because synaptic failures, depression, or facilitation can alter average EPSP size relative to measured EPSP sizes. Furthermore, measurements are variable, for example $J_E$ for layer 4 to layer 4 connections in rodent barrel cortex has been estimated to be 1.6 mV {\em in vitro} \citep{Feldmeyer_etal99} vs.\ 0.66 mV {\em in vivo} \citep{Schoonover_etal14}. If we assume typical values for $J_E$ of $0.5-1$~mV, then $1.5 J_EK_En_E$ would exceed 75~mV for $\sqrt{Kn}>7-10$ and exceed 150~mV for $\sqrt{Kn}>10-14$ (compare values of $\sqrt{Kn}$ in Table~\ref{spines}). We can very roughly guess that neural responses may become better described by tight rather than loose balance somewhere in this range of mean  excitatory input (and corresponding $\sqrt{Kn}$). While the measurements of excitatory currents and conductances described above argue that such a range is not reached in primary sensory cortex, it could conceivably  be reached (Table~\ref{spines}) in areas with higher $K_E$, \eg\ frontal cortex.

\subsubsection*{Nonlinear Behaviors}

Sensory cortical neuronal responses display a variety of  nonlinear behaviors that, as we'll describe, are expected from the SSN loosely balanced regime but not from the tightly balanced regime. Many of these nonlinearities are often summarized as "normalization" \citep{Carandini_Heeger12}, meaning that responses can be fit by a phenomenological model of an unnormalized response that is divided by (normalized by) some function of all of the unnormalized responses of all the neurons within some region. To describe these nonlinear behaviors, we must first define the classical receptive field (CRF): the localized region of sensory space in which appropriate stimuli can drive a neuron's response. 

One nonlinear property is sublinear response summation: across many cortical areas, the response to two stimuli simultaneously presented in the CRF is less than the sum of the responses to the individual stimuli, and is often closer to the average than the sum of the individual responses \citep[reviewed in][]{Reynolds_Chelazzi04,Carandini_Heeger12}. An additional nonlinearity is that the form of the summation changes with the strength of the stimulus: summation becomes linear for weaker stimuli \citep{Heuer_Britten02,Ohshiro_etal13}. It is often difficult to determine if such nonlinear behaviors are computed in the recorded area or involve changes in the inputs to that area. However, some recent experiments studied summation of response to an optogenetic and a visual stimulus, a case in which the inputs driven by each stimulus should not alter those driven by the other. Sublinear summation of responses to these stimuli was found (\citealp{Nassi_etal15,Wang_etal19}, but see \citealp{Histed18}), which became linear for weak stimuli \citep{Wang_etal19}.

Another set of nonlinearities involve interaction of a CRF stimulus and a ``surround" stimulus, which is located outside the CRF. Across many cortical areas, surround stimuli can suppress response to a CRF stimulus (``surround suppression"; reviewed in \citealp{Rubin_etal15,Angelucci_etal17}), but this effect varies with stimulus strength. When the center stimulus is weak, a surround stimulus can facilitate rather than suppress response  \citep{Sengpiel_etal97,Polat_etal98,Ichida_etal07,Schwabe_etal10,Sato_etal14}. Similarly, the {\em summation field size} -- the size of a stimulus that elicits strongest response, before further expansion yields surround suppression -- is largest for weak stimuli and shrinks with increasing stimulus strength \citep{Anderson_etal01,Song_Li08,Sceniak_etal99,Cavanaugh_etal02a,Shushruth_etal09,Nienborg_etal13,Tsui_Pack11}. The summation field size in feature space -- the optimal range of simultaneously presented motion directions in monkey area MT -- similarly shrinks with increasing stimulus strength \citep{Liu_etal18}.

Additional nonlinearities include a decrease, with increasing stimulus strength, in the ratio of excitation to inhibition received by neurons \citep{Adesnik17} and in the wavelength of a characteristic spatial oscillation of activity \citep{Rubin_etal15}.

All of these nonlinear cortical response properties, and more, follow naturally  \citep{Ahmadian_etal13,Rubin_etal15} from the two regimes of the loosely balanced scenario with a supralinear input/output function, along with simple assumptions on connectivity (\eg\ that connections decrease in strength and/or probability with spatial distance, \eg\ \citealp{Markov_etal11}, or with difference in preferred features, \eg\ \citealp{Ko_etal11,Cossell_etal15}). In contrast, as described previously, the tightly balanced scenario causes population-averaged responses to be linear responses to the input (individual neurons, but not the population average, may have nonlinear behaviors), and thus appears inconsistent with these nonlinear cortical behaviors, which in most cases are consistent enough across neurons that they should characterize the mean population response. Synaptic nonlinearities can give nonlinear population-averaged behavior in the tightly balanced regime \citep{Mongillo_etal12}, but it has not been claimed or demonstrated that this could produce specific nonlinearities like those seen in cortical responses.

\subsubsection*{Correlations and Variability}
Across many cortical systems, the correlated component of neuronal variability is decreased when a stimulus is given, with variability decrease seen both in neurons that respond to the stimulus and those that don't respond \citep{Churchland_etal10}. This is also naturally explained by the loosely balanced SSN network \citep{Hennequin_etal18}. In the strongly coupled regime of the loosely balanced SSN network, increasing stimulus strength increases the strength with which correlated patterns of activity inhibit themselves, thus damping their responses to input fluctuations. The tightly balanced state represents the end state of this process -- a fully asynchronous regime in which correlations are completely suppressed (\citealp{vanVreeswijk_Sompolinsky98,Renart_etal10}; with dense connectivity, the mean correlation is proportional to $1/\Kn$, and the standard deviation of the distribution of correlations is proportional to $1/\sqrt{\Kn}$ \citet{Renart_etal10}; recall that $\Kn$ is meant to be a very large number to achieve the tightly balanced state). Thus, the tightly balanced state appears incompatible with the observed decrease in correlated variability induced by a stimulus, because the state has essentially no correlated variability. However, it should be noted that variants of the tightly balanced network involving structured connectivity can yield finite correlated variability among preferentially connected neurons while maintaining tight balance, although average correlation over all neuron pairs can still go to zero with increasing $\sqrt{\Kn}$ \citep{Rosenbaum_etal17,Litwin-Kumar_etal12}.

\subsection*{Discussion}

We have seen that many independent lines of evidence are all consistent with cortex being in a loosely balanced regime, and are inconsistent with tight balance. We define balance to mean that the dynamics yields a systematic cancellation of excitation by inhibition; a signature of this for the loosely balanced scenario that we consider is that the net input a neuron receives after cancellation grows sublinearly as a function of its external input. Loose balance means that the net input after cancellation is comparable in size to the factors that cancel, whereas tight balance means that the net input is very small relative to the cancelling factors. In both cases, the net input after cancellation is comparable in size to the distance from rest to threshold so that neuronal firing can be in the fluctuation-driven regime that produces irregular firing like that observed in cortex. 

One line of evidence for loose balance involves a variety of measurements on the numbers and/or strengths of the inputs cells receive, including spine counts, strengths of external and total input, and strengths of excitatory and of inhibitory input. These measurements show that the expected ratio of mean to standard deviation of the network input before any tight balancing is already consistent with the ratios observed for a cell's net input as judged by its voltage response. That is, tight cancellation is not needed to achieve the ratios observed. These measurements further show that external input and network input are comparable in size to the net input remaining after cancellation, and that they and the total excitatory and total inhibitory input are all comparable to the distance from rest to threshold, consistent with loose but not tight balancing. Other lines of evidence include a variety of nonlinear population response properties of sensory  cortical neurons, as well as the presence of correlated variability in neural responses and its decrease upon presentation of a stimulus, all of which emerge naturally from loose balance with a supralinear input/output function, but appear largely incompatible with tight balance.

It should be emphasized that the number of excitatory synapses received by an excitatory cell, $K_E$, increases from primary sensory to higher sensory to frontal cortex \citep[\eg][]{Elston03}. Higher numbers are expected to push in the direction of tighter balance.  The  expected ratio of input mean to standard deviation and the expected size of the mean input both can become high enough to potentially yield tight balance for the highest $K_E$'s, particularly if higher average firing rates $r_E$ are assumed. Our other arguments depend largely, but not entirely, on measurements from sensory cortex. 
The measurements of net input and external input are all from primary sensory cortex. The studied nonlinear response properties are primarily from both lower and higher visual cortices (reviewed in \citealp{Rubin_etal15}). Suppression of correlated variability by a stimulus, however, has been observed in frontal and parietal as well as sensory cortex \citep{Churchland_etal10}. In sum, while the evidence strongly favors loose balance in sensory cortex, the evidence as to the regime of frontal cortex is weaker.

The seminal discovery of the tightly balanced network \citep{vanVreeswijk_Sompolinsky96,vanVreeswijk_Sompolinsky98} solved a key problem in theoretical neuroscience: how can neurons remain in the fluctuation-driven regime, so that they have irregular firing with reasonable firing rates, without requiring fine tuning of parameters?  The answer was that when external and network inputs were very large, the network's dynamics could robustly tightly balance the excitation and inhibition that neurons receive, leaving a net input after cancellation that is negligibly small relative to the factors that cancel. This allows both the mean and standard deviation of the net input to be comparable to the distance from rest to threshold despite the very large size assumed for the factors that cancel, yielding the fluctuation-driven regime. This achievement along with the model's mathematical tractability have made it a popular model for the theoretical study of neural circuits. However, for all of the reasons stated above, this tightly balanced regime does not seem to match observations of at least sensory cortical anatomy and physiology. 

The loosely balanced solution shows that, when neuronal input/output functions are supralinear, the same dynamical balancing can arise from network dynamics without fine tuning, but in a regime in which external and network inputs are not large. Instead, the balancing arises when these inputs, and the net input remaining after cancellation of excitatory and inhibitory input, are all comparable in size to one another and to the distance from rest to threshold. Furthermore, for weak inputs this same scenario produces a weakly-coupled, feedforward-driven regime which explains the observation that summation changes from sublinear, or suppressive, for stronger stimuli to linear, or facilitative, for weak inputs.

The tightly balanced network demonstrated that a network could self-consistently generate its own variability. As we showed in the section ``How large is $\sqrt{Kn}?"$, the loosely balanced regime can also generate realistic levels of variability. However, biologically there is no need for the network to generate all of its own variability, as all inputs to cortex are noisy (and there are other sources of noise such as stochasticity of cellular and
synaptic mechanisms \citep{Mainen_Sejnowski95,Schneidman_etal98,Odonnell_vanRossum14} and input correlations \citep{Stevens_Zador98,DeWeese_Zador06}). In at least one case \citep{Sadagopan_Ferster12}, the noise derived from the cortical area's input was shown to be large enough to potentially fully account for the noise seen in the cortical neurons. In the SSN network, the network will amplify input noise in the weakly coupled regime, and then decrease noise for increasingly strong inputs in the strongly coupled, loosely balanced regime; the result is that, for higher input strengths, noise can be reduced to the level driven by the inputs \citep[\eg, see Fig.~2D of][]{Hennequin_etal18}, consistent with the observations of   \citet{Sadagopan_Ferster12}.

In conclusion, we believe that at least sensory, and perhaps all of, cortex operates in a regime in which the inhibition and excitation neurons receive are loosely balanced. This along with the supralinear input/output function of individual neurons and simple assumptions on connectivity explains a large set of cortical response properties. A key outstanding question is the computational function or functions of this loosely balanced state and the response properties it creates (\eg, see \citealp{Echeveste_etal19}; G. Barello and Y. Ahmadian, in preparation).

\subsection*{Appendix 1: Nomenclature for balanced solutions}

There is no standard nomenclature for describing balanced solutions. Here we have used loose vs.\ tight balance to describe, given systematic cancellation, whether the remainder after cancellation is comparable to, or much smaller than, the factors that cancel.

\citet{Deneve_Machens16} used loose balance to mean that fast fluctuations of excitation and inhibition are uncorrelated, although they are balanced in the mean, as in the sparsely-connected network of \citet{vanVreeswijk_Sompolinsky98}; and used tight balance to mean that fast fluctuations of excitation and inhibition are tightly correlated with a small temporal offset, as in the densely-connected network of \citet{Renart_etal10} and in the spiking networks of Deneve, Machens and colleagues in which recurrent connectivity has been optimized for efficient coding \citep{Boerlin_etal13,Bourdoukan_etal12,Barrett_etal13}. All of these networks are tightly balanced under our definition. 

\citet{Hennequin_etal17} also defined balance to be tight if it occurs on fast timescales, and loose otherwise, but they implied that this is equivalent to our definition, that tight balance means the remainder is small compared to the factors that cancel, and loose balance means the remainder is comparable to the factors that cancel.
The implied equivalence rests on the fact that tight balance under our definition produces very large (\ie, $O(\sqrt{K})$) negative eigenvalues (in linearization about the fixed point) which means very fast dynamics, approaching instantaneous population response as $K$ and hence negative eigenvalues go to infinity.
 We point out, however, that loose balance under our definition can produce negative eigenvalues large enough to produce quite fast dynamics, with effective time constants on the order of single neurons' membrane time-constant, or even as small as a few milliseconds, depending on parameters.
 
An additional source of confusion is that there are two forms of fast fluctuations, with different behaviors.  Fast fluctuations can be 
shared (correlated) across neurons, or they can be independent. The large negative eigenvalues in tightly balanced networks (in our definition) affect shared fluctuations corresponding to eigen-modes in which the activities of excitatory and inhibitory neurons fluctuate coherently. Thus, shared fluctuations are balanced on fast time scales. By contrast, spatial activity patterns in which neurons fluctuate independently are largely unaffected by those eigenvalues, and need not be balanced. 

Fluctuations due to changes in population mean rates of the external input are shared, and so this form of fluctuation is followed on fast time scales in all balanced networks (at finite rates in loosely balanced networks, and approaching instantaneous following in tightly balanced networks). Fluctuations also arise
from network and external neuronal spiking noise. In networks with sparse connectivity \citep{vanVreeswijk_Sompolinsky98}, this yields independent fluctuations in different neurons, and thus independent fluctuations of excitation and inhibition on fast time scales (though their means are balanced). In networks with dense connectivity \citep{Renart_etal10}, these spiking fluctuations become shared fluctuations
due to common inputs arising from the dense  connectivity, and so in these networks excitation and inhibition are balanced on fast time scales. To summarize, all balanced networks will balance shared fluctuations, such as those due to changing external input rates, on fast time scales; but excitation and inhibition can nonetheless be unbalanced on fast time scales in sparse networks, due to independent fluctuations induced by spiking noise.

To conclude, we would argue for a future standardized  terminology for dynamically-induced balancing of excitation and inhibition, in which ``loose" vs.\ ``tight" balance  refer to our definition as to whether the remainder after cancellation is comparable to, or much smaller than, the factors that cancel. We suggest the use of ``temporal" vs.\ ``mean" balance to refer to whether or not excitation and inhibition are balanced on fast time scales, which depends on whether there are substantial shared input fluctuations across neurons. ``Finite" vs.\ ``instantaneous" time scales of balancing can distinguish whether relaxation rates -- the rates of balancing shared fluctuations -- are moderately-sized vs.\ very large.

\subsection*{Appendix 2: When do balanced solutions arise?}

We consider a rate model in which the neuron's input/output function is described by some function $f(x)$, which is zero for $x\leq 0$ and monotonically increasing for $x\geq 0$. Then the network's steady-state  firing rate $\rss$ for a steady input $\vI$ is
\be
\rss = f(\vW\rss+\vI)
\label{ssgeneral}
\ee
where $f$ acts element by element on its argument, that is, $f(\vmu)$ is a vector whose  $i^{th}$ element is $f(\mu_i)$ (the $f$'s might differ for different neurons, which we neglect for simplicity). As before, we let $\psi=\|\vW\|$ and $c= \|\vI\|$. We define the dimensionless and $\om(1)$ matrix $\vJ=\vW/\psi$ and vector $\vg=\vI/c$, so that $\vJ$ and $\vg$ represent the relative synaptic strengths and relative input strengths, respectively, while their overall magnitudes and dimensions are in $\psi$ and $c$. Then, as in \citet{Ahmadian_etal13}, we can define the dimensionless variable $\vy=\frac{\psi}{c}\vr$, and Eq.~\ref{ssgeneral} can be rewritten
\be
\yss=  \frac{\psi}{c}f\left(c(\vJ\yss+\vg)\right)
\label{ssdimless0}
\ee
Note that this equation ensures that $\yss\geq 0$. Note also that, when $f(x)=(x)_+^\pow$ ($(x)_+=x$, $x\geq 0$; $=0$, otherwise), then this equation can be rewritten $\yss=  \alpha\left(\vJ\yss+\vg\right)_+^\pow$ where $\alpha=\psi c^{\pow-1}$. This is the origin of the dimensionless constant $\alpha$ mentioned in the main text.

If we define $f^{-1}(0)=0$, then because $f$
is monotonically increasing for non-negative arguments, it is invertible over that range, \ie\ $f^{-1}(x)$ is defined for $x\geq 0$. We can then rewrite Eq.~\ref{ssdimless0} as 
\be
\frac{1}{c}\,f^{-1}\left(\frac{c}{\psi}\yss\right)=(\vJ\yss+\vg)_+
\label{rectified}
\ee
If we assume 
\be
(\vJ\yss+\vg)_i\geq 0 \text{ for all } i,
\label{condition}
\ee
that is, if $(\vJ\yss)_i\geq -(\vg)_i$ for all $i$, then
we can replace the right side of Eq.~\ref{rectified} with $\vJ\yss+\vg$ (without the $()_+$). This condition, Eq.~\ref{condition}, is a condition on the solution $\yss$, which we must check is self-consistently met for any solution we derive under this assumption. Note also that, from Eq.~\ref{ssdimless0}, the condition ($\vJ\yss+\vg)_i> 0$ is met if and only if $(\yss)_i>0$, so if we find a solution $\yss$ that has all positive elements, it will automatically satisfy Eq.~\ref{condition}. Given this assumption, a bit of further manipulation then yields
\be
\yss= -\vJ^{-1}\vg+ \frac{1}{c}\,\vJ^{-1}f^{-1}\left(\frac{c}{\psi}\yss\right)
\label{ssdimless}
\ee
The first term,  $\yss^0\equiv -\vJ^{-1}\vg$, is the balancing term, which cancels the external input $\vg$, \ie\ $(\vJ\yss^0 + \vg)=0$.  If  the second term becomes small relative to the first in some limit, then the tightly balanced solution, $\yss\approx -\vJ^{-1}\vg$ or equivalently $\rss \approx -\vW^{-1}\vI$, exists in that limit, while a loosely balanced solution (balance index $\om(1)$) arises when the 2nd term is comparable in size to the first. (More careful analysis is needed to ensure that this solution is stable, and that there are not also other solutions.) Note that Eq.~\ref{ssdimless} gives an equation of the form Eq.~\ref{loose} when we (1) Take $f^{-1}(x)=(x)_+^{1/\pow}$ and
(2) Multiply both sides of Eq.~\ref{ssdimless} by $c/\psi$ to convert $\yss$ to $\rss$.

Assuming all the elements of $\yss^0\equiv -\vJ^{-1}\vg$ are $> 0$, a self-consistent solution in which the second term in Eq.~\ref{ssdimless} becomes small can be found in at least three cases: 
\begin{itemize}
\item If $c$ and $\psi$ are scaled by the same factor, which becomes arbitrarily large, then there is a self-consistent solution in which $\yss$ is converging to $-\vJ^{-1}\vg$. Then
the $f^{-1}$ factor is not changing (except for the small changes due to the changes in $\yss$ as it converges), but it is multiplied by the factor $\frac{1}{c}$, which becomes arbitrarily small; thus the second term becomes arbitrarily small, regardless of the particular structure of $f$. This is the case studied for the tightly balanced solution, where both $c$ and $\psi$ are taken proportional to $\sqrt{K}$ with $K$ very large. (Note that the mean field equations derived in \cite{vanVreeswijk_Sompolinsky98} differ from the generic steady-state rate equations,  \req{ssgeneral}, in that they also involve the self-consistently calculated input fluctuation strengths, $\sigma_A$; the scaling argument given here nevertheless holds in that case too.)

\item Suppose $c$ is scaled for fixed $\psi$, which is the biological case in which synaptic strengths are fixed and the strength of the external input is varied from small to large. Then if $f^{-1}(x)$ grows more slowly than linearly in increasing $x$, then the $\frac{1}{c}$ factor shrinks faster than the $f^{-1}$ term grows, so again there is a self-consistent solution in which $\yss$ is converging to $-\vJ^{-1}\vg$ and the second term becomes arbitrarily small with increasing $c$. This is the case studied for the loosely balanced solution in the SSN, in which $f(x)$ grows supralinearly with $x$ and therefore $f^{-1}(x)$ grows sublinearly with $x$.
\item We again suppose $c$ is scaled for fixed $\psi$, but now imagine that $f^{-1}(x)$ grows faster than linearly in increasing $x$, \ie\ $f(x)$ is sublinear (for example, $f(x)=(x)_+^\pow$ for $0<\pow<1$). Then there is a self-consistent solution in which $\vy\rightarrow -\vJ^{-1}\vg$ as $c\rightarrow 0$, with the second term in Eq.~\ref{ssdimless} going to zero as $c\rightarrow 0$. This case is the reverse of the SSN: the strongly coupled, balanced regime arises for $c\rightarrow 0$, while the weakly coupled, feedforward-driven regime arises for large $c$.
\end{itemize}
In sum, if the elements of $-\vJ^{-1}\vg$ are positive, then a self-consistent tightly-balanced solution arises for any $f$ if $c$ and $\psi$ are scaled together by an increasing factor; for supralinear $f$ if $c$ is scaled by an increasing factor; or for sublinear $f$ if $c$ is scaled by a decreasing factor. In all of these cases, for moderate sizes of the scaled parameter(s) (\eg, for the SSN, for $\alpha=\om(1)$) such that the second term of Eq.~\ref{ssdimless} is comparable in size to the first, a loosely balanced solution should arise. Note that, since $\rss=\frac{c}{\psi}\yss$, then from Eq.~\ref{ssdimless} the net input after balancing should grow with increasing external input $c$ as $f^{-1}(c)$; this is sublinear in $c$ for the SSN case of supralinear $f$.

If one considers a two-population model -- a population of E cells and a population of I cells, with each population's average rate  represented by a single variable -- then conditions on $\vJ$ and $\vg$ can be defined such that the elements of $-\vJ^{-1}\vg$ are positive and the balanced fixed point is stable and is the only fixed point \citep{vanVreeswijk_Sompolinsky98,Ahmadian_etal13,Kraynyukova_Tchumatchenko18}. On the other hand, when the E element of $-\vJ^{-1}\vg$ is negative, Eq.~\ref{ssdimless} cannot serve as a basis for an asymptotic expansion with the leading term $-\vJ^{-1}\vg$, and the tightly balanced state does not exist. (Given $(-\vJ^{-1}\vg)_{\msub{E}}<0$, if the tightly balanced state existed -- meaning that the second term of Eq.~\ref{ssdimless} becomes much smaller than the first while Eq.~\ref{ssdimless} is applicable -- then $\yss$ must have crossed zero to become negative, but once that has happened we could no longer proceed past Eq.~\ref{rectified} and Eq.~\ref{ssdimless} would no longer be applicable, which is a contradiction; hence the tightly balanced state cannot exist.)
However, the loosely balanced state can still arise in this case in a broad parameter regime of $\vJ$ and $\vg$, and can be found as the fixed point of the iterative equation, $\yss(t)= -\vJ^{-1}\vg+ \frac{1}{c}\,\vJ^{-1}f^{-1}\left(\frac{c}{\psi}\yss(t-1)\right)$, given appropriate initial conditions $\yss(0)$; see \citet{Ahmadian_etal13}. In this case, with increasing $c$, $\re$ grows, but  then saturates and starts decreasing, and eventually is pushed down to 0. However, if we assume that the maximal external input (\ie\ the maximal $c$; for example, the maximal firing rate of the thalamic input to a primary sensory cortical area) can only drive $\re$ to saturation or slightly beyond, this represents a viable model of cortical systems \citep{Ahmadian_etal13,Rubin_etal15,Hennequin_etal18}.

A two-population model accurately describes the behavior of an unstructured model with many E and I neurons, \ie\ with random connectivity and with neurons in each population receiving comparable stimulus inputs. In some cases this model also can form a good approximation to the behavior of a multi-neuron circuit with structured connectivity and stimulus selectivity \citep{Ahmadian_etal13}. More generally, though, in such a structured circuit with localized connectivity, for larger/stronger localized stimuli, some set of neurons (\eg, neurons not selective for the stimulus) may eventually  receive a net inhibition and become silent, meaning that the condition of Eq.~\ref{condition} is not met and Eq.~\ref{ssdimless} does not apply. (However, if the connectivity is translation-invariant -- the same at any position in the model -- and the external input extends more narrowly than the network connections, then a balanced fixed point can still be attained, \citealp{rosenbaum2014balanced}.) Nonetheless, we find in simulations \citep{Ahmadian_etal13,Rubin_etal15,Hennequin_etal18} that for reasonable stimulus input strengths, SSN behavior is reasonably described by the two-population model, in that (1) there is a transition with increasing input strength from a weakly coupled, feedforward-driven regime to a strongly-coupled, loosely balanced regime in which the input to excited neurons grows sublinearly as a function of the external input strength; and (2) if we define the $\vW$ and $\vg$ of the two-population model as describing the net input received by a cell in the larger, structured model -- \eg, $\wee$ represents the mean summed synaptic strength from excitatory cells to a single excitatory cell, and $\gge$ represents the mean external input received by stimulus-selective excitatory cells -- then reasonably good insight into the operating regime of the larger model can be obtained from the analysis of the two-population model presented here and, in much more detail, in \citet{Ahmadian_etal13,Kraynyukova_Tchumatchenko18}. 

We believe the same overall analysis of a transition from a weakly coupled regime to a strongly coupled, loosely balanced regime will apply to multi-population models incorporating multiple subtypes of inhibitory cells \citep[\eg][]{Litwin-Kumar_etal16,Kuchibhotla_etal17,GarciaDelMolino_etal17}, but more detailed aspects of the analysis of the two-population model \citep{Ahmadian_etal13,Kraynyukova_Tchumatchenko18} need to be generalized to that case.

\subsection*{Acknowledgements} We thank Larry Abbott, Mario DiPoppa and Agostina Palmigiano for many helpful comments on the manuscript, David Hansel, Gianluigi Mongillo and Alfonso Renart for many useful discussions, and Randy Bruno for help with references. YA is supported by start-up funds from the University of Oregon. KDM is supported by NSF DBI-1707398, NIH U01NS108683, NIH R01EY029999, NIH U19NS107613, Simons Foundation award 543017, and the Gatsby Charitable Foundation.

\bibliography{base,new}
\end{multicols}
\end{document}